\documentclass[prb,a4paper,twocolumn,showpacs,preprintnumbers,amsmath,amssymb,floatfix]{revtex4}

\usepackage{graphicx}
\usepackage{dcolumn}
\usepackage{bm}

\newcommand{\Slowlevel}[1]{(TMTSF)$_2$#1}
\newcommand{\SX}{\Slowlevel{$X$}}
\newcommand{\NO}{\Slowlevel{NO$_3$}}
\newcommand{\PF}{\Slowlevel{PF$_6$}}

\newcommand{\TNO}{(TMTTF)$_2$NO$_3$}

\newcommand{\myreffig}[1]{Fig.\ \ref{#1}}
\newcommand{\myrefeq}[1]{Eq.\ (\ref{#1})}

\DeclareMathOperator{\sech}{sech}
\newcommand{\mdeg}{^\circ}
\newcommand{\mTSDW}{T_C}

\newcommand{\mv}[1]{\mathbf{#1}}               

\newcommand{\mva}{\mv{a}}  \newcommand{\va}{$\mva$}
\newcommand{\mvb}{\mv{b}}  
\newcommand{\mvbp}{\mv{b'}}  
\newcommand{\mvc}{\mv{c^\star}}   
\newcommand{\mvk}{\mv{k}}   
\newcommand{\mvq}{\mv{q}}   
\newcommand{\mvB}{\mv{B}}  

\newcommand{\plane}[2]{#1\text{--}\,#2}
\newcommand{\mbcplane}{\plane{\mvbp}{\mvc}}
\newcommand{\bcplane}{$\mbcplane$}

\newcommand{\imag}{\mathrm{i}}
\DeclareMathOperator{\realpart}{Re}

\begin{document}


\title{Hidden symmetries in Bechgaard salt \NO}

\author{Kazumi Maki}
    \affiliation{Department of Physics and Astronomy, University of Southern California, Los
Angeles CA 90089-0484, USA\\ and\\ Max Planck Institute for the
Physics of Complex Systems, N\"{o}thnitzer Str.\ 38, D-01187,
Dresden, Germany}
    \email{kmaki@usc.edu}

\author{Mario Basleti\'{c}}
    \affiliation{Department of Physics, Faculty of Science, P.O.Box 331, HR-10002 Zagreb, Croatia}

\author{Bojana Korin-Hamzi\'{c}}
    \affiliation{Institute of Physics, P.O.Box 304, HR-10001 Zagreb, Croatia}

\author{Silvia Tomi\'{c}}
    \affiliation{Institute of Physics, P.O.Box 304, HR-10001 Zagreb, Croatia}
\date{\today}

\begin{abstract}
Among known Bechgaard and Fabre salts \NO\ is unique since it never becomes superconducting even under pressure. Also, though \NO\ undergoes the spin
density wave (SDW) transition, the low temperature transport is semimetallic and gapless. We propose: a) the absence of the superconductivity is due to the inverse symmetry breaking associated with the anion ordering at $45\,$K; b) the SDW state below $9\,$K should be unconventional as seen from the angle dependent magnetoresistance oscillation (AMRO); c) a new phase diagram for Bechgaard salts where unconventional spin density wave (USDW) occupies the prominent space.
\end{abstract}


\maketitle

\section{Introduction}
The Bechgaard salt \PF\ is well known as the first organic superconductor discovered by Jerome {\it et al}\cite{JeromeJP80} in 1980. Further, Bechgaard salts [\SX\ with $X$$=$PF$_6$, ClO$_4$, AsF$_6$\ldots] and Fabre salts (TMTTF)$_2$$X$ are well known with a variety of ground states under pressure and in magnetic field: spin density wave (SDW), unconventional spin density wave (USDW), field induced spin density wave (FISDW) and triplet superconductivity.\cite{IshiguroBook98,ChaikinSHEGOLEV,BasleticPRB02,KangSM03,DoraEPL04,BasleticPRB07,LeeJPSJ06}

\NO\ (and possibly \TNO) is unique among known Bechgaard and Fabre salts\cite{MoriJPSJ06} in that it never becomes superconducting even under high pressure.\cite{KangPRL90} Further, \NO\ undergoes the SDW transition at $9\,$K, but its low temperature transport is semimetallic and gapless.\cite{BasleticPRB07}

As described in section \ref{ch2} the absence of the triplet superconductivity is due to the inversion symmetry breaking associated with the anion ordering at $45\,$K.\cite{AndersonPRB84,WonPREPARATION} The absence of the superconductivity is most conveniently formulated in terms of the chiral symmetry breaking term or the Rashba term.\cite{RashbaSPSS60} The Rashba term of the order $2-3\,$K is adequate to suppress the triplet superconductivity completely. Further, the partial suppression of FISDW in \NO\ (i.e., the absence of a FISDW for $P < 8\,$kbar and $B < 20\,$T) observed in Ref.\ \onlinecite{VignollesPRB05} gives the Rashba term $\sim 8\,$K.

In section \ref{ch3} we briefly review AMRO in \NO. In our analysis the Landau quantization of the quasiparticle spectrum discovered by Nersesyan {\it et al} \cite{NersesyanJLTP89,NersesyanJPCM91} plays the crucial role. Indeed, we have succeeded in interpreting the surprising angle dependent magnetoresistance oscillation (AMRO)\cite{BasleticEPL93,BiskupPRB94} and the nonlinear Hall resistance\cite{BasleticSSC96} in terms of USDW.\cite{BasleticPRB07}

Finally, we propose a new phase diagram for Bechgaard salts\cite{BasleticPRB07} which is very different from the earlier version, for example in Ref.\ \onlinecite{IshiguroBook98}. In the new phase diagram, USDW will occupy a wider region in the metallic side.

\section{Role of the inversion symmetry breaking\label{ch2}}
In 1984.\ P.\ W.\ Anderson\cite{AndersonPRB84} suggested that the triplet superconductivity is forbidden in crystals without inversion symmetry. Then the discovery of a new superconductivity in CePt$_3$Si, the crystal without inversion symmetry, by Baner {\it et al} in 2004 has generated sensation.\cite{BauerPRL04,FrigeriPRL04,FrigeriNJP04,IzawaPRL05,MakiPB06}

Fortunately, the situation is much simpler in \NO\ since we expect that the Rashba term should be of the order of $10\,$K as it is associated with the AO at $45\,$K Also, the anion ordering is along the \va\ axis with $\mvq = (1/2,0,0)$. So, we take the Rashba term as:
\begin{equation}
    H_R = \lambda \sigma_x \hat{k}_x \cos\left(\frac{1}{2}\mva\mvk\right)
\end{equation}
as the simplest choice. Then the effect of the Rashba term in the superconducting transition temperature $T_C$ is given by\cite{WonPREPARATION}
\begin{equation}
    -\ln\frac{T_C}{T_{C0}} = \left\langle|f|^2\right\rangle^{-1}
        \left\langle |f|^2 \left\{
            \realpart \psi\left(\frac{1}{2} + \imag\frac{\lambda\hat{k}_x\cos(\mva\mvk/2)}{2\pi T_C} \right)
            -\psi\left(\frac{1}{2}\right)\right\}\right\rangle
\end{equation}
where $T_{C0}$ is the superconducting transition temperature in the absence of the Rashba term (e.g.\ $T_{C0} = 1.2\,$K) and $\psi(z)$ is the digamma function. Here $\langle\cdots\rangle$ means the average over the Fermi surface and $f$ is the one of candidate function of chiral $f$-wave superconductors.\cite{DominguezCondMat0601065} The average over the Fermi surface is readily done and we find:
\begin{equation}
\label{eq3}
    -\ln\frac{T_C}{T_{C0}} =
            \realpart \psi\left(\frac{1}{2} + \imag\frac{\lambda_0}{2\pi T_C} \right)
            -\psi\left(\frac{1}{2}\right)
\end{equation}
where $\lambda_0=\lambda\cos(\mva\mvk_F/2) \simeq \lambda\cos(\pi/8)$. We note that the \myrefeq{eq3} is the same as the expression for $T_C$ in the presence of the Pauli term.\cite{SarmaJPCS63,MakiPTP64} In particular the superconductivity is completely suppressed for $\lambda_0 \ge 2.5\,$K. Actually, the partial suppression of FISDW as found in Ref.\ \onlinecite{VignollesPRB05}, provides the more precise value for $\lambda_0$.\cite{WonPREPARATION} The suppression of FISDW is given by a similar formula as \myrefeq{eq3} where $T_C$ and $T_{C0}$ have to be replaced by the corresponding FISDW transition temperatures. Then comparison with the experimental data in Ref.\ \onlinecite{VignollesPRB05} gives $\lambda_0 \sim 7.5\,$K.

Therefore \NO\ appears to be the simplest case where the Rashba term on the triplet superconductivity can be studied. Also we predict that \TNO\ neither becomes superconducting even under high pressure.

\section{AMRO in \NO\label{ch3}}
Unconventional density wave (UDW) is a kind of density wave where the quasiparticle energy gap $\Delta(\mvk)$ vanishes on lines on the Fermi surface.\cite{DoraMPL04} For example in UDW the electric resistance is metallic and in many cases $R \propto T^2$. The quasiparticle energy spectrum in the absence of magnetic field is given by:
\begin{equation}\label{eq4}
E^{\pm}(\mvk) = \eta(\mvk) \pm \sqrt{\xi^2+\Delta^2(\mvk)}
\end{equation}
where $\eta(\mvk)$ is the imperfect nesting, $\xi = v(k-k_F)$ with $v$ the Fermi velocity, and $\Delta(\mvk)=\Delta\cos(\mvb\mvk)$ for \NO.

\begin{figure}
\includegraphics*[width=8.6cm]{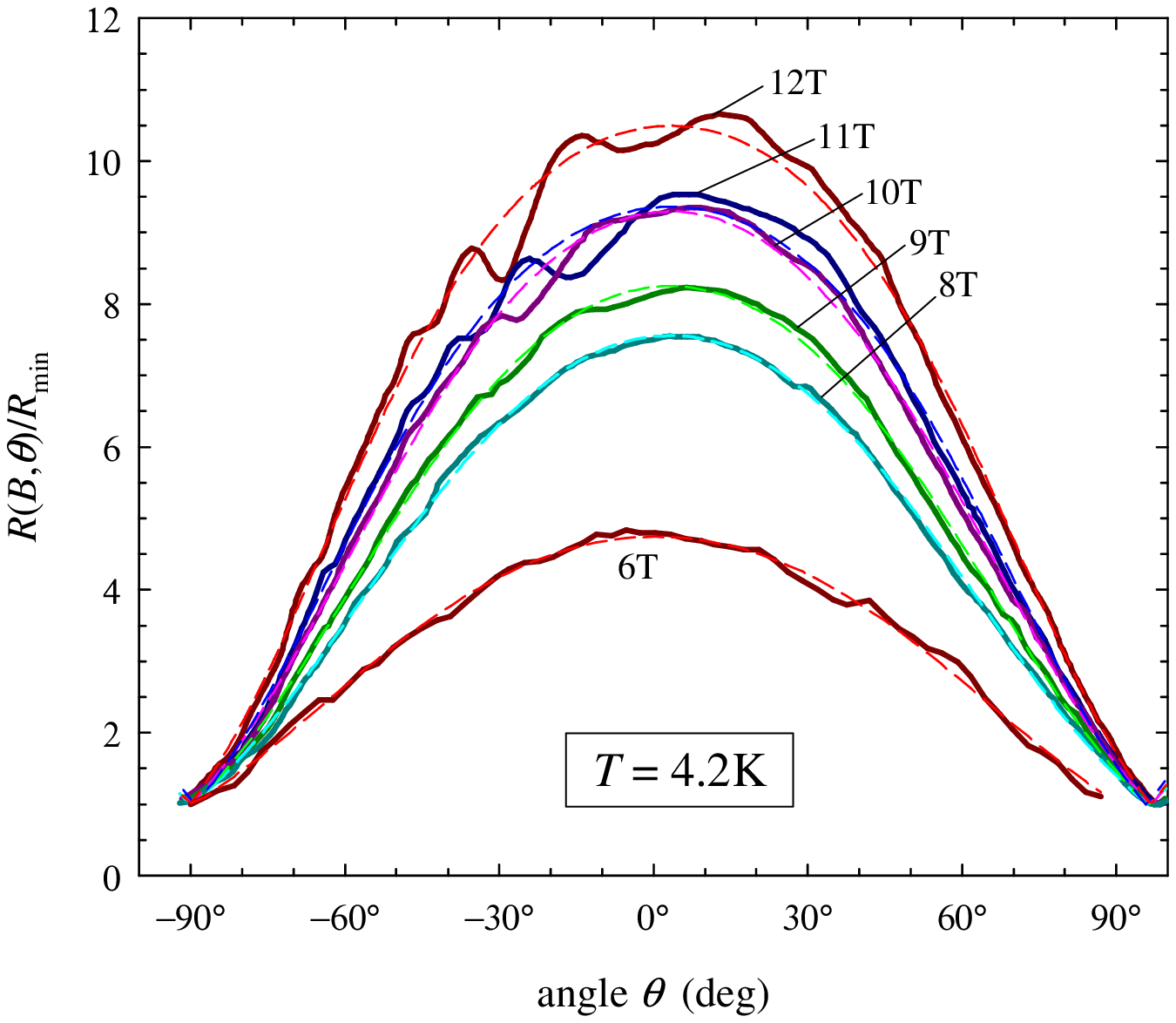}
\caption{\label{fig1}The angular dependence of the normalized
resistance $R(B,\theta)/R_0$ at $T = 4.2\,$K (full lines:
experimental data; dashed lines: fits to the theory). Magnetic field
is rotated in \bcplane\ plane, and $\theta=0\mdeg$ corresponds to
$\mv{B}\|\mvc$. Data are from Ref.\ \onlinecite{BiskupPRB94}
($B\geq8\,$T) and Ref.\ \onlinecite{BasleticEPL93} ($B=6\,$T).}
\end{figure}

Then in the presence of a magnetic field $\mvB$ tilted form the $\mvc$ axis by an angle $\theta$, the quasiparticle energy changes to
\begin{equation}\label{eq5}
E_n^{\pm} = \eta_n \pm \sqrt{2neB|\cos\theta|vb\Delta}
\end{equation}
with $n = 0,1,2,3,\ldots$ due to the Landau quantization.\cite{NersesyanJLTP89,NersesyanJPCM91} Then the magnetoresistance is given by
\begin{equation}\label{eq6}
R_{xx}^{-1} = \sigma_1\left[
    1 + 2C_1\sech^2(x_1/2) + \cdots
\right]
\end{equation}
where $x_1 = E_1^+/k_BT$ and $C_1$ is a constant weakly dependent on $T$ and $B$. In \myreffig{fig1} we show a fitting of the AMRO for different magnetic fields, from Ref.\ \onlinecite{BasleticEPL93} and \onlinecite{BiskupPRB94}.

In this fitting we have neglected $\eta_1$. The inclusion of $\eta_1$ should reproduce the bumpy structure in $R_{xx}$. With this fitting we obtain $\Delta = 6.3\,$K Compared with $\mTSDW = 9\,$K this $\Delta$ appears to be somewhat too small. But it is possible that the Rashba term would suppress $\Delta$ to a small value. The pressure dependencies of $\mTSDW$ and $\Delta$ are therefore of great interest.

\section{New phase diagram\label{ch4}}
\begin{figure}
\includegraphics*[width=8.6cm]{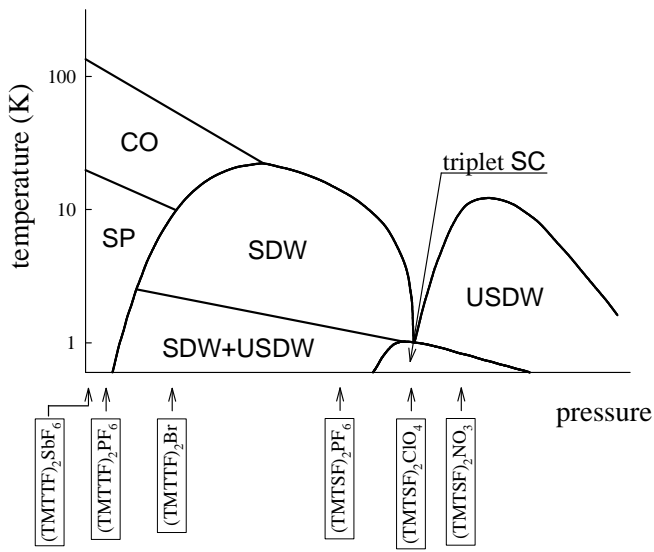} 
\caption{\label{fig2}The schematic pressure--temperature
phase diagram for Bechgaard salts. (`\textsf{CO}' = charge order, `\textsf{SP}' = spin Peirerls.)
}
\end{figure}

In the preceding section we have seen that SDW in \NO\ is USDW.\cite{BasleticPRB07} From this we can construct a new phase diagram for Bechgaard salts. This is shown in \myreffig{fig2}.
Compared with the old one in Ref.\ \onlinecite{IshiguroBook98} and Ref.\ \onlinecite{MoriJPSJ06}, we have added a) the coexistence region of SDW plus USDW for $T<T^\star \sim \mTSDW/3$ and b) USDW for $P \gtrsim 8\,$kbar in \PF. Furthermore, charge order (CO) in (TMTTF)$_2$SbF$_6$ and the $T^\star(=T_c/3)$ anomaly in (TMTTF)$_2$Br under pressure have been reported in Refs.\ \onlinecite{NadJPF05,NagasawaJPF05,NomuraJPF05,NadJPSJ06,TakahashiJPSJ06,NomuraSM05}. From this phase diagram the absence of the triplet superconductivity in \NO\ is very surprising, as we have clarified in section \ref{ch2}.

\begin{acknowledgments}
We thank Balazs D\'{o}ra, Stephen Haas, Hyekyung Won and Pierre Monceau for helpful collaborations and discussions. MB, BKM and ST acknowledge support from the Croatian Ministry of Science, Education and Sports projects No. 119-1191458-1023 and 035-0000000-2836.
\end{acknowledgments}

\bibliography{MakiNewZealand}

\end{document}